# WearWrite: Orchestrating the Crowd to Complete Complex Tasks from Wearables

# (We Wrote This Paper on a Watch)

Michael Nebeling[1], Anhong Guo[1], Kyle Murray[3*],
Annika Tostengard[3], Angelos Giannopoulos[3] Martin Mihajlov[3],
Steven Dow[1], Jaime Teevan[2], Jeffrey P. Bigham[1]

| [1]Human-Computer Interaction Institute | [2]Microsoft Research | [3]Freelancer |
| Carnegie Mellon University | Redmond, WA | Recruited on oDesk.com |
| [mnebelin, anhongg, spdow, jbigham]@cs.cmu.edu | teevan@microsoft.com | * helped with development not writing |

*This 6-page paper was written using the WearWrite system for Android Wear smartwatches. WearWrite lets an author direct a crowd to complete the complex task of writing a paper from their smartwatch. The author injects necessary expertise, while the crowd does the actual writing. This document contains three versions of the paper in sequence: first, the final version edited on a desktop by the authors; second, the draft created by the crowd as orchestrated by an author from his watch; and finally, an initial outline entered by the first author using WearWrite.*

**ABSTRACT**
In this paper we introduce a paradigm for completing complex tasks from wearable devices by leveraging crowdsourcing, and demonstrate its validity for academic writing. We explore this paradigm using a collaborative authoring system, called WearWrite, which is designed to enable authors and crowd workers to work together using an Android smartwatch and Google Docs to produce academic papers, including this one. WearWrite allows expert authors who do not have access to large devices to contribute bits of expertise and big picture direction from their watch, while freeing them of the obligation of integrating their contributions into the overall document. Crowd workers on desktop computers actually write the document. We used this approach to write several simple papers, and found it was effective at producing reasonable drafts. However, the workers often needed more structure and the authors more context. WearWrite addresses these issues by focusing workers on specific tasks and providing select context to authors on the watch. We demonstrate the system's feasibility by writing this paper using it.

**INTRODUCTION**
Smartwatches like the Android Smartwatch, Apple Watch, and Samsung Gear Live make it possible for people to remain connected wherever they are from easy-to-access devices on their wrists. Wearable devices are quickly advancing from being merely entertaining to being capable of performing a wide range of important tasks. However, performing complex tasks is difficult on a wearable device because the interfaces provided by wearable technology are limited. Text input is hard from a watch, and very little content can be displayed at any one time. Smartwatches are often designed to be used when the user's attention is fragmented, with the user being able to pay attention to the wearable device only in micro moments lasting a few seconds at most. These constraints make it challenging to create user experiences that retain the full range of functionality available with large screens and keyboards.

We believe it is possible to support complex watch-based tasks despite these limitations. Complex computing tasks may be difficult to perform wholly using wearable devices due to their constrained interaction capabilities, but they can be successfully completed from wearable devices by directing the crowd in how to do them on one's behalf. A domain expert using only a wearable device is able to 1) contribute bits of necessary expertise and 2) manage the contributions of crowd workers, so that the task is performed similarly to what the expert could achieve on a desktop computer. Tasks that cannot currently be performed using wearable technology alone can be completed in this hybrid manner.

The specific task we explore is academic writing. This is a challenging domain to crowdsource because the technical content of an academic paper requires expertise to explain, and writing a good paper requires maintain global context. We present *WearWrite*, a system that allows authors who do not have access to large devices to contribute unique insight related to their domain of expertise and big picture direction from their watch, while using crowd workers who have access to additional context to implement the changes suggested by the author from their desktop computers and work ensure consistency. Enabling academic authors to contribute their unique expertise to a project through wearable technology allows them to take advantage of free micromoments. Additionally, by employing crowd workers to create content and only requiring the author to respond to it, WearWrite takes advantage of the fact that it is easier to read text on a wearable device (even on the tiny screen) compared to writing new text.

In order to explore solutions to the limitations presented by smartwatch wearable technology, we conducted a series of experiments to test whether papers could be written from a smartwatch using a simple interface. The work presented here tests whether the skills of crowdsource workers are capable of producing high quality academic papers, the effectiveness of the application WearWrite, and offer insights for future work in crowdsourcing and wearable technology. The contributions of this paper are as follows:

- We introduce and motivate a new paradigm for combining expertise from users from wearables computing used to orchestrate a crowdsourcing-based workflow for performing knowledge-intensive tasks
- We validate this approach in the case of academic writing by writing this (and several other) papers.
- We offer a number of insights into the feasibility of the WearWrite approach for working on complex problems within the constraints of a smartwatch.

**RELATED WORK**

Research related to the topics discussed here includes crowdsourcing, collaborative writing, and wearable technology. Crowdsourcing has been used to complete complex tasks, including writing. Soylent [2] splits writing projects into stages and invites crowd workers to make suggestions and to shorten and proofread text. For the management of the crowd work, CrowdForge [7] identifies broken flows between complex tasks and manages the project by filling the dependencies between them. The idea of "shepherding the crowd" [5] was introduced to help give workers feedback so they could improve over time. Ensemble [6] uses a team leader to direct writing projects. The complementary writing skills of individuals produce better results in less time and with higher creativity. Finally, Flash Teams has explored the idea of bringing together on-demand teams of experts for specific tasks, which can be applied across a wide variety of domains, including writing [9]. WearWrite represents a different approach to crowdsourced writing in which expertise is made available and constrained by the availability and limitations of a wearable device.

Wearable devices have recently received a great deal of attention in the literature. A number of projects have been introduced to both increase the input capabilities of smartwatches and to increase the amount of information that can be shown to a smartwatch user [1,11] Despite this, interaction on wearable devices like watches remains quite limited. Text input has been explored [8] but is much slower than from other types of devices; an author would not want to write an entire academic paper this way. Speech recognition is the current standard for input to smartwatches, but it can be error prone, especially for long sequences of text. WearWrite overcomes existing limitations by integrating wearable input with input from other types of devices.

**WEARWRITE**

We now describe the WearWrite system. The system was developed iteratively. After discussing our experiences with the first few attempts and sharing what we learned, we present the details of the final system.

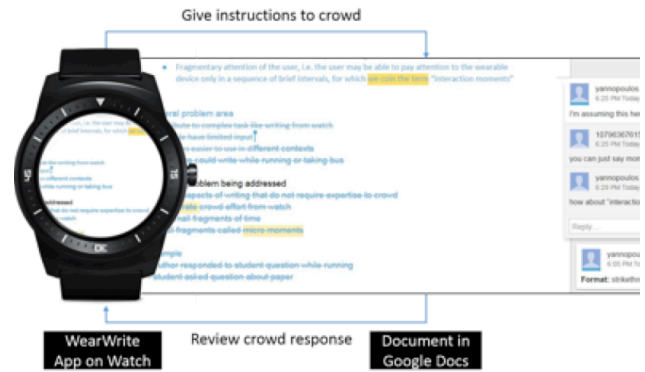

**Figure 1:** WearWrite allows watch users to send high-level instructions to the crowd. In turn, the crowd

**Iterative Design**

To develop the WearWrite system, we iteratively explored two psuedo-watch-based experiences for writing academic content. Each experiments assessed the viability of the design. The experiments involved hiring four to five hourly writers to work on a document shared using Google Docs (drive.google.com). Google Docs allows any number of workers to contribute to a document at the same time and also includes a commenting system. Each experiment asked workers to write a paper. In these preliminary experiments, the researchers did not provide the initial instructions, feedback, or comments from the watch. However, their interactions with workers were constrained to limitations imposed by watch interactions.

The first preliminary trial involved writing a 2-page paper based on data collected by the authors, that was eventually titled *A Survey of Shortening Tasks in Crowdsourcing Markets*. To collect the data in the paper, the authors hired crowd workers using Amazon's Mechanical Turk, asked them to shorten a snippet of text, and asked a few follow up survey questions. The data were compiled into a spreadsheet and given, along with a few bullets highlighting key findings, to workers hired using the freelancing platform oDesk.com. Workers were asked to use this content to write a research paper, creating the necessary graphics to support the content. The authors approved or rejected each change. While this method produced high quality local content, it produced limited framing content. Worker writing quality varied as did the quality and type of images provided by oDesk hires, and workers wanted more structure. The nature of Google Doc's editing system resulted in an overwhelming number of edits that were difficult for the authors to approve or reject individually.

The second trial attempted to duplicate an existing published research paper, titled *Removed for Anonymity*. Workers from oDesk.com were given a document template and bulleted suggestions for content to include in each section of the report, as well as the images used in the existing research paper. In this trial, workers frequently commented on each other's work, which alleviated some of the organizational effort required by the authors in the first trial. While collaboration between workers was high, the notifications system in Google Docs lacked necessary context for the authors to provide comments and feedback efficiently. One worker found the original source.

We used what we learned from these two trials to implement the WearWrite system. The first draft of this research paper was written by workers hired on oDesk.com with work overseen by the research team from smartwatches. Our experiences with this process are discussed after a detailed description of the system.

### Watch Interaction

WearWrite enables a smartwatch user to orchestrate crowd workers by supporting editorial interactions. First, it supports direct but limited authoring of new content via speech recognition. Authors can add sections, paragraphs, and bullets. Second, as crowd workers edit the document, their changes are displayed on the smartwatch where the expert can approve or reject them. Edits are sent along with a screenshot of the entire page in which an edit was made. Finally, comments written by crowd workers are sent to the smartwatch and the author can reply using speech. The whole document can be browsed in a thumbnail representation.

### Architecture and Implementation

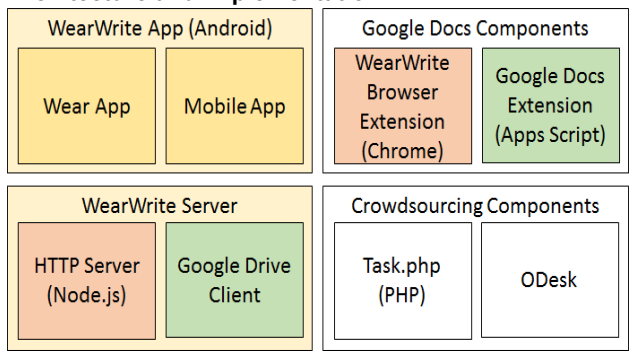

**Figure 2:** The four main components of the WearWrite system, which allow crowd workers on desktop interfaces and an author on a smartwatch to collaborate.

WearWrite's architecture can be broken down into four major interconnected blocks. On the author side, a WearWrite Android application communicated with an Android Wear smartwatch and also the WearWrite Server. The WearWrite server connects to Google Docs to make changes, notice changes or comments, and facilitate discussion. Crowd workers are recruited from oDesk, and coordinated using a task queue that contains a number of generic paper-writing tasks, e.g., "turn the bullets in the 'Introduction' into paragraphs," "find the paper that each bullet in the 'Related Work' section refers to and create a reference to it."

1. **WearWrite App**: an Android Mobile app and an Android Wear app running on the smartwatch.
2. **WearWrite Browser Extension**: used for taking screenshots, extracting edits/comments from the document, and triggering the insertion of new edits via Google Docs Extension and new comments via Google Drive Client into the document[1]
3. **Google Docs Extension**: used for analysing document structure of, and insertion of new edits into, the document.
4. **WearWrite Server**: 1) HTTP server used by a) the WearWrite App for posting new edits/comments from the watch to the document and b) the WearWrite Browser Extension for uploading screenshots of the document; and 2) Google Drive Client used for insertion of new comments into the document.
5. **Task.php**: directs workers to specific tasks to provide lightweight structure.
6. **ODesk**: used for recruiting workers

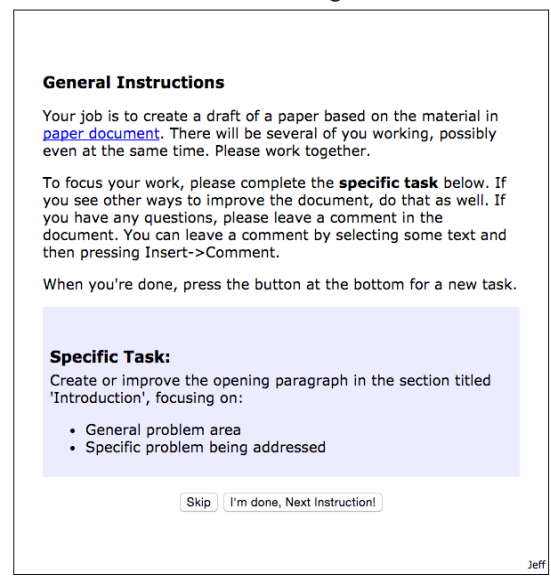

**Figure 3: An example of the task.php task queue.**

---

[1] currently, comments can only be inserted via the Google

**WRITING THIS PAPER**

We now turn to our experience writing this paper with WearWrite. Our original intention was to write *everything* on the watch, but this proved unwieldy. In practice, we expect authors will switch between the watch, a mobile, and a desktop computer, as is convenient for them.

In the ideal scenario, the watch user performs all interactions using exclusively the smartwatch. The watch user can receive general-purpose emails on the smartwatch, but exclusively uses the WearWrite prototype, running on the smartwatch, for all crowd sourcing and document writing activities. The workflow followed is:

1. The watch user receives a request for information through general-purpose email on the smartwatch.
2. The watch user recruits a number of crowd workers through the smartwatch. In applications where crowd workers are available on short notice, it is assumed that a preparatory recruitment process has already been performed using a desktop computer, and that in this scenario the watch user communicates using the smartwatch with crowd workers who are on standby and recruits those who are available when needed.
3. After sending requests to crowd workers, the watch user efficiently uses time spent waiting for replies to seed the document. The seed document is created with Google Docs and authored exclusively from the smartwatch. It contains the basic structure of the document, the key points it should address, and references for crowd workers to use.
4. When crowd workers become available, an iterative authoring process begins. The crowd workers transform the seed document into the desired final document, step by step. They work on concurrently through Google Docs, and every edit they make is handled as a suggestion. The suggestion is sent to the watch user to accept or reject, and potentially comment on. Also, any comments made by crowd writers are sent to the watch user, who can reply to them if necessary. The watch user can also view the entire document on the smartwatch, although this is not easy on the small watch display. At the end of this process, the document draft is complete.

In practice, we compromised in this a bit:

1. Crowd workers were recruited from the oDesk desktop computer web interface.
2. Overcoming the occasional prototype bugs was achieved by using a desktop computer if necessary. However, the desktop was only used to perform actions that would normally have been performed using a bug-free prototype.
3. The watch user spent nearly all of his time using the smartwatch, not a desktop computer.
4. On a few occasions, crowd workers emailed the watch user. In these cases, they were asked to repeat their message through the Google Doc, where they were responded to using WearWrite.
5. Two different researchers adopted the role of watch user at various times.
6. Finally, in the actual experiment, five crowd workers were recruited

The practical scenario is equivalent to the ideal scenario for the purposes of this experiment because the collaborative authoring process was executed entirely through the watch interface as intended; rarely using a desktop computer to overcome a prototype bug, but even then only performing actions as the prototype was designed to do. The recruiting process is an important part of the practical crowdsourcing activity, but it is not a part of the collaborative authoring process which is the focus of the experiment. Similarly, sharing the watch user role between researchers did not impact the interface usage or the collaborative workflow.

**EXPERIENCES WRITING A PAPER ON A WATCH**

In this section, we report on writing this paper.

The interactions between crowd workers and WearWrite users were systematically observed while writing this paper and during the iterative design process. Our observations and conclusions can be classified as related to *a)* preparation and task allocation, *b)* basic effectiveness, *c)* communication, *d)* difficulties, *e)* involvement and creativity.

*Preparation and task allocation*

The first interactions between the crowd workers and the watch user were through a third party website, oDesk.com. The normal process of recruitment concluded with crowd workers receiving their initial instructions. Once they started work on the collaborative authoring job itself, the crowd workers exhibited a constructive pattern of behaviour. They started by working through the entire seed document, in order to understand the research they were documenting. Next, they worked on tasks allocated to them by *task.php*. Randomized task allocation allowed multiple crowd workers to collaborate on the authoring job without prior coordination. However, some problems occurred. For instance, a worker was asked to write the paper's abstract before the rest of the paper had been composed. As the crowd workers became more accustomed to the collaboration dynamic in the experiment, they took increasingly greater initiative in simply choosing the section of the paper they felt they were best able to contribute to.

*Basic effectiveness*

Workers were, in general, effective at producing content. The entire paper was written within three days. Readers can judge for themselves, but the authors were surprised by the quality of the content produced. The authors had significant "idle time" during the authoring process. The author could see how the document progressed, and it was sufficient to occasionally provide the crowd workers with suggestions and feedback on edits.

The crowd workers appeared comfortable working from the original bullet points comprising the seed document. They used these bullet lists both as content expressing the paper's ideas, and as a to-do list from which they could work. They progressed through the work systematically, and marked ongoing and completed work for each other (for example, using a strikethrough effect on completed bullet points, but not deleting them).

Furthermore, the workers performed intelligent and critical work. Beyond their basic task of transforming notes into well-structured text, they made suggestions and improvements ranging from re-wording keywords and key phrases in the notes to planning and executing ideas for the presentation. They made suggestions about the structure of the document as well as comments that were valuable at the research level, such as how best to evaluate the experimental results. The more creative contributions of the crowd workers are discussed below.

*Communication*
The crowd workers approached their task as a highly collaborative process. They regularly made comments to and requested feedback from the author. They used document comments to request clarifications from the author if they felt the instructions, the seed document, or another crowd worker's text to be unclear. In turn, sometimes text or comments written by the crowd workers were vague, in which case the watch user initiated a discussion in the document comments. The discussion was rapid, concrete and constructive, as evidenced by the fact that the entire document was completed within three days.

Although communication through document comments was performed successfully, some crowd workers were unsatisfied with the approach, suggesting, for example, in-document "chat" which is not supported through the watch. Future work could certainly provide useful improvements by offering a better discussion interface.

*Difficulties*
Workers experienced a bit of a learning curve before they became comfortable with WearWrite. One crowd worker, following the randomised task assignment process of task.php started to write the abstract before working on the rest of the document, and missed many of the key points. The watch user intervened by making a comment on the half-written abstract, requesting that it be deferred until later. Authors may sometimes start with abstracts, but this is difficult for workers new to the project to do without necessary context.

In general, linear progression by crowd workers through the document is not a good idea. The task allocation process was altered for this final version to make clear that workers could skip its suggestions. Beyond this, moving from fully randomised task allocation to a state where crowd workers are comfortable enough to choose tasks for themselves is advantageous, since crowd workers will perform better in this case. However, some randomisation and/or centralised control of the task allocation process needs to be maintained to avoid a case where crowd workers "pick low hanging fruit" and certain writing tasks are left unperformed because nobody volunteered for them.

Additional difficulties include maintaining a cohesive style and terminology (expert, domain expert, watch user, smartwatch user, main author were used interchangeably by the various crowd workers – as we do this final editing pass we are trying to converge on vocabulary, but future systems may try to get workers to agree on these terms before contributing too much text), managing the length of the document (indeed, this document grew to be too long), and complicated change management, which was not needed for this experiment.

Some writing tasks that would be easy for the author can be unexpectedly difficult for workers who lack specific expertise and context. For example, the "Related Work" section ended up the poorest of all of the document, despite the author having provided projects to reference and snippets describing their importance. Workers are not experts in the field, and it is difficult to write critically about unfamiliar research.

Finally, the author faced certain difficulties using WearWrite. The speech recognition on the watch produced transcription errors that were sometimes confusing. Also, there were some editing bugs that were hard to explain during the experiment; on one occasion, the author attempted to accept an edit made by a crowd worker but the edit was deleted instead. However, the fact that the paper was successfully authored in a brief period of time using only the intended interfaces indicates these difficulties are relatively minor.

*Involvement and creativity*
The crowd workers took ownership of the writing task and demonstrated significant amounts of involvement and creativity. There was a suggestion for an alternative scenario for the paper's proposed paradigm, although we found it to be too complex. Feedback was given on terminology defined in the seed document, which was in some cases accepted and improved the document's clarity, for example "interaction moments" replaced the seed document's term "mico-moments", or the term "watch metaphor", which was discarded.

The crowd workers also made long-term contributions. They communicated their reflections on task design and pointed out potentially problematic issues with the current prototype, for example that the speech recognition used on the smartwatch can introduce confusion by incorrectly transcribing the WearWrite user's comments. They made suggestions about how the current experiment might be more extensively assessed, for example by tracking document metrics over time/effort. They provided ideas for future research, for example integrating voice interactions with document comments based interactions.

At the same time, the crowd workers were sufficiently detached from the writing task to be able to think as peer-reviewers for the paper. This is in fact also a creative contribution, as it often starts a discussion that leads to improvements of the paper. For example, it was suggested that we clearly explain why the topic for the text collaboratively authored during the experiment was representative.

**DISCUSSION**

Together, results indicate that "complex tasks that are difficult to perform on wearable devices due to their constrained interaction capabilities can still be successfully crowdsourced from wearable devices." WearWrite enables its users to "contribute the necessary expertise" and to "manage the contributions of crowd workers," allowing the user to author a research paper using a wearable device and collaborating with crowd workers. With this system, one achieves results comparable to those from a desktop computer.

This approach might be useful for completing other types of complex tasks or for completing such tasks on other types of wearable devices, such as head mounted displays. It remains an open question for future research and philosophical consideration to address questions of the influence of this paradigm on cognitive processes. How would allowing work to invade these small moments, which might otherwise be rest from intellectual work, affect psychological health? The potential negative influence of constant "uptime" performing intellectually challenging tasks may be severe.

WearWrite breaks authoring tasks into small steps. Some steps are performed by the expert, while others are outsourced to crowd workers. Even the tasks performed by the expert are performed in small "interaction moments." However, at what point does this seeming benefit become a liability, if for instance, by splitting our attention among multiple unrelated authorings, necessary context instantiation is lost? As a counterargument, is it perhaps the case that by outsourcing the less creative parts of a project, intellectually intensive work is reserved for the situations in which it is actually merited.

Breaking up and carefully orchestrating the authoring process grants an opportunity to study the process itself. For example, the collaborative writing process could offer insight on the learning process in order to help students improve their writing.

Finally, we must also discuss the role of crowd workers in this paradigm. Should crowd workers be considered authors of papers written in this way? In this context, it should be noted that many co-authorship scenarios exist where collaborators are not, in practice, given credit: for example, grammar editors, ghostwriters, and technical writers, etc. We have listed the worker who contributed to this paper as authors of it. The contributions of the crowd workers are harder to encapsulate in a role as simple and atomic as these. As we mentioned earlier, workers often participated creatively in the writing process.

**CONCLUSION**

We have introduced WearWrite, an application for Android Wear that allows authors to write academic papers from their smartwatches. We believe this system represents a new paradigm that may change how and where we work together to get work done. Completing complex tasks within the constraints of wearable devices brings up a host of interesting questions regarding the impact of wearable devices on the way work is performed, and on the people who perform it that we have only started to address.

*The following is the draft of the paper as it existed at the end of the WearWrite process. Four workers were hired and produced this draft from the initial outline directed by the watch user. We accepted whatever pay rate the workers listed on their profile, which was from $5.56 to $50.00 per hour (we ended up paying all workers at least $10/hour). Workers spent 36 hours and 20 minutes working on the draft, which cost $810 USD. Although somewhat expensive, it represented nearly a full work week, and resulted in a draft that was a good starting point for final revisions.*

*In total, workers submitted 95 revisions starting from the initial outline. In these revisions, the crowd workers made 117 insertions, 170 replacements, 107 deletions, and 48 formatting changes. After the first day of the experiment, the watch user began inserting meta comments on observations of the process thus far. During the entire experiment, the watch user made 68 comments answering questions by the crowd and giving additional instructions as required.*

*We asked each worker if they would like to be an author of the paper after they had a chance to see what they had produced. Three of the 4 not only agreed, but said they would be happy to continue working on it for free in that case, e.g.,*

> *"Thank you for the consideration. I will gladly accept :) Please let me know if I can help further with any aspects. My contribution would be off-the clock."*

*To facilitate comparison, the crowd's text is shown in the lefthand column of this section, whereas the result after our final editing pass is shown in the righthand column. The final version is much shorter than the crowd's version.*

# WearWrite: Orchestrating the Crowd to Complete Complex Tasks from Wearables

# (We Wrote Most of This Paper On A Watch)

**ABSTRACT**

This paper introduces and validates a paradigm for overcoming the interface limitations of wearable technologies by leveraging crowdsourcing. The introduced paradigm enables the application of wearable technologies to tasks for which their interface limitations would normally render them unusable. The solution proposed leverages the capabilities of the crowd for performing tasks that cannot be performed using the wearable device. This approach is valuable because it allows experts who do not have access to a larger device to contribute their expertise, while freeing them of the obligation to perform simpler parts of the overall task. The solution enables the expert, using the wearable device, to manage the workflow followed by the crowd workers, as well as provide them with expert knowledge. Specifically, this paper presents an experiment in which an expert uses an android smartwatch to guide five crowd workers in the task of collaboratively authoring an academic paper in the expert's domain of expertise. The smartwatch runs a prototype of the new, custom software WearWrite. The collaborative authoring process is achieved using Google Docs and a WearWrite Google Docs extension which implements the communication channel connecting the expert to the crowd workers. The experimental results are analyzed and demonstrate that the experiment validated our general paradigm.

**ABSTRACT**

In this paper we introduce a paradigm for completing complex tasks from wearable devices by leveraging crowdsourcing, and demonstrate its validity for academic writing. We explore this paradigm using a collaborative authoring system, called WearWrite, which is designed to enable authors and crowd workers to work together using an Android smartwatch and Google Docs to produce academic papers, including this one. WearWrite allows expert authors who do not have access to large devices to contribute bits of expertise and big picture direction from their watch, while freeing them of the obligation of integrating their contributions into the overall document. Crowd workers on desktop computers actually write the document. We used this approach to write several simple papers, and found it was effective at producing reasonable drafts. However, the workers often needed more structure and the authors more context. WearWrite addresses these issues by focusing workers on specific tasks and providing select context to authors on the watch. We demonstrate the system's feasibility by writing this paper using it.

# INTRODUCTION

Modern wearable technology affords access to technical capabilities that are very similar to desktop computing capabilities and the new wearable devices produced have fast advanced from merely entertaining to being capable of performing a wide range of useful tasks. . Already, smartwatches like the Apple Watch and Samsung Gear Live facilitate tasks that were initially possible only from other forms of computerised devices with larger user interfaces such as the smartphones, tablets and other PCs. Today, nearly all of these capabilities are replicated on smartwatches and these devices can now be used for increasingly complex tasks.

These useful functions have their share of challenges, though, and there are two major differences between running applications on a traditional computerized device and those running on wearable technology. First, the interfaces provided by wearable technology are different, and in some ways very limited. Users are unable to type fast in response to emergency situations, since only a single hand can type on the tiny smartwatch screen. Typographical errors will likely convey unintended meanings, and issues with delayed data loading time witnessed with the Apple watch may also delay response. These factors impede group communication, a necessary factor for accomplishing complex tasks. Second, the actual context in which wearable technology is useful may be limiting. For instance, the portability and size of smartwatch means that a professional task could be completed while jogging, and so user interfaces designed for such devices must accommodate a wide variety of activities and contexts. The challenge is to allow users of wearable technology to perform tasks that normally require access to desktop computing. While the computing capabilities are similar, it is challenging to create user interfaces that retain the full range of function available with large screens and keyboards.

This research focuses on tasks that can not be fully performed using wearable technology only. Identified were a class of applications in which user interface requirements exceed the capabilities of today's wearable technology and scarce application-domain expertise is required. There is a very high value in enabling appropriate experts to contribute their expertise through wearable technology, since the reduction in workload resulting from applications such as WearWrite will make it easier for experts to contribute to a greater number of projects. WearWrite's goal is to allow the experts to crowdsource the work that they are unable to perform using their wearable device(s) and orchestrate crowd efforts from a smartwatch using short interactions called micro-moments. This means that wearable technology must allow them to perform knowledge-intensive work that cannot be crowdsourced and manage the crowdsourcing process.

# INTRODUCTION

Smartwatches like the Android Smartwatch, Apple Watch, and Samsung Gear Live make it possible for people to remain connected wherever they are from easy-to-access devices on their wrists. Wearable devices are quickly advancing from being merely entertaining to being capable of performing a wide range of important tasks. However, performing complex tasks is difficult on a wearable device because the interfaces provided by wearable technology are limited. Text input is hard from a watch, and very little content can be displayed at any one time. Smartwatches are often designed to be used when the user's attention is fragmented, with the user being able to pay attention to the wearable device only in micro moments lasting a few seconds at most. These constraints make it challenging to create user experiences that retain the full range of functionality available with large screens and keyboards.

We believe it is possible to support complex watch-based tasks despite these limitations. Complex computing tasks may be difficult to perform wholly using wearable devices due to their constrained interaction capabilities, but they can be successfully completed from wearable devices by directing the crowd in how to do them on one's behalf. A domain expert using only a wearable device is able to 1) contribute bits of necessary expertise and 2) manage the contributions of crowd workers, so that the task is performed similarly to what the expert could achieve on a desktop computer. Tasks that cannot currently be performed using wearable technology alone can be completed in this hybrid manner.

At this point, it is useful to consider an example scenario. A research team resolves to replicate an experiment documented in a previously published paper. They solicit certain clarifications from the author of the paper, however their funding constraints are such that they can not wait for these clarifications. The author of the paper receives the request and realises that making the clarifications is in fact of critical importance. However, she is currently on a sailing holiday, without access to a laptop or tablet. She has a smartwatch, whose interface is, however, insufficient for performing all the background research and authoring tasks necessary for providing a good response. Furthermore, she does not want to provide mere thoughts, ideas and pointers as a response, as this would be unprofessional. Rather, she needs to author a white paper, which will add new analysis to the discussion of her original paper, and publish it on her institutional website. The solution is to crowd-source the work that requires a desktop computing interface, while herself injecting expertise where necessary.

This particular research focuses on smartwatch devices. Smartwatches are very interesting, cutting edge and perhaps still slightly controversial.Nevertheless, they are accessible enough to make them easy to experiment with. These devices exemplify some key constraints of wearable computing such as small screen size. The distinguishing interaction features addressed are:

- Relative ease of reading text on the wearable device (even on the tiny screen) compared to writing new text (where location may affect speech recognition)
- Fragmentary attention of the user, i.e. the user may be able to pay attention to the wearable device only in a sequence of brief intervals, called "interaction moments"

In order to explore solutions to the limitations presented by smartwatch wearable technology, a series of experiments were conducted to test whether papers could be written from a smartwatch using a simple interface. The task of authoring an academic paper represents a sufficient test of whether complex tasks could successfully be completed by allowing experts to inject knowledge where necessary and manage the contributions of crowd workers. The work presented here tests whether the skills of crowdsource workers are capable of producing high quality academic papers, the effectiveness of the application WearWrite, and offer insights for future work in crowdsourcing and wearable technology.

Useful computing tasks, which cannot be wholly performed using wearable devices due to their constrained interaction capabilities, can be successfully partially crowdsourced. A domain expert using only a wearable device is able to 1) contribute the necessary expertise and 2) manage the contributions of crowd workers, so that the task is performed similarly to what the expert could achieve on a desktop computer. We consider the task of

The specific task we explore is academic writing. This is a challenging domain to crowdsource because the technical content of an academic paper requires expertise to explain, and writing a good paper requires maintain global context. We present *WearWrite*, a system that allows authors who do not have access to large devices to contribute unique insight related to their domain of expertise and big picture direction from their watch, while using crowd workers who have access to additional context to implement the changes suggested by the author from their desktop computers and work ensure consistency. Enabling academic authors to contribute their unique expertise to a project through wearable technology allows them to take advantage of free micromoments. Additionally, by employing crowd workers to create content and only requiring the author to respond to it, WearWrite takes advantage of the fact that it is easier to read text on a wearable device (even on the tiny screen) compared to writing new text.

In order to explore solutions to the limitations presented by smartwatch wearable technology, we conducted a series of experiments to test whether papers could be written from a smartwatch using a simple interface. The work presented here tests whether the skills of crowdsource workers are capable of producing high quality academic papers, the effectiveness of the application WearWrite, and offer insights for future work in crowdsourcing and wearable technology. The contributions of this paper are as follows:

- We introduce and motivate a new paradigm for combining expertise from users from wearables computing used to orchestrate a crowdsourcing-based workflow for performing knowledge-intensive tasks
- We validate this approach in the case of academic writing by writing this (and several other) papers.
- We offer a number of insights into the feasibility of the WearWrite approach for working on complex problems within the constraints of a smartwatch.

authoring an academic paper to be a representative task and a sufficient test of this paradigm.

In summary, the contribution of this paper comprises the following:

- Introduced and motivated a new paradigm of wearable computing used to orchestrate a crowdsourcing-based workflow for performing knowledge-intensive tasks
- Developed new software, based on the new paradigm, catering to the specific case of writing a research paper, as a collaboration between main authors using exclusively wearable devices, and crowd worker collaborators
- Demonstrated a successful case study, by using the new software in practice
- Formally validated our central thesis in our main experiment, using the new software
- Analysed our results so as to offer insights for future work in this area

**RELATED WORK**
Research related to the topics discussed here includes crowdsourcing, collaborative writing and wearable technology. . A joint study by the University of California in San Francisco examined the use of the SmartWatch device in effectively detecting abnormal motion patterns that are usually typical of GTC seizures and found that the watch was able to decipher the difference between normal motions and GTC seizures.(Sullivan, 2013). These wearables worked by sending text messages and/or phone calls through Bluetooth links to computers or smartphones within the user's proximity. The computer or smartphone would be on the user or within the same room as the user of the watch. The incident was then recorded along with the time, date as well as the duration of the wrist movements. The wrist movement would be interpreted as an event of convulsion by the SmartWatch.

Fifteen patients were recorded by the SmartWatches as having had a total of seven GTC seizures while they were being monitored. All the seven seizure incidences were identified by the SmartWatches and signal validated by EEG/video. Several crowdsourcing markets have been developed to facilitate the completion of large tasks and implement a system for easy communication and feedback. Soylent (Bernstein et al., 2010) splits writing projects into stages and invites crowd workers to make suggestions and to shorten and proofread text. For the management of the crowd work, CrowdForge (Kittur et al., 2011) identifies broken flows between complex tasks and manages the project by filling the dependencies between them. Ensemble (Kim et. al., 2014) uses a team leader to direct writing projects. The complementary writing skills of individuals produce better results in less time and achieves maximum creativity. In all of the above approaches, management is provided by the division of large projects into tasks and then coordinating the tasks. These platforms do not rely purely on automatic writing systems and relies on crowdsourcing for real time interactions so that, regardless of how many tasks are required to complete a large project, automated errors in grammar, spelling and syntax are always corrected using human efforts..

In the case of wearable technology, automatically generated responses may fail to logically respond to questions and comments and would have to track and contextualize each comment in order to give coherent answers. To address this, VizWiz, a visual answering software for iPhone, answers questions in real time with responses generated by humans hired from Amazon's Mechanical Turk (Bigham et al., 2010). VizWiz uses the qTurkit approach and aims to consistently maintain a pool of workers so that someone is always available to answer questions. WearWrite builds on the concept of VizWiz, but reduces time by expanding the pool to workers from oDesk.com.

WearWrite is inspired by growing number of apps indeed gaining crowd based applications in the corporate world platform, health and security sectors in particular. Police out in the field may use the Google Glass and related eyewear technology to gain situational awareness and transmit the collected data to their remote datacenter. The same interconnectivity has been used by doctors situated at different remote locations to share critical patient information. In such contexts where timely data transmission is imperative at the least possible time is a great perk, the predictive user response options provided by the Apple Watch replaces the need to type the whole text from the watch before sending.

In a similar fashion, a sociometric badge was created (in replacement of the normal organizational tags) that had the capability to record common human activities, extract features of speech in real time, perform indoor user localization through triangulation with reduced estimation errors, communicate with bluetooth-enabled phones and use IR sensors to record time spent in face-to-face interactions. As a result, this wearable badge collected data from different employees within the subject organization and relayed the data in real time for survey of different aspects of organizational behavior, the punchline of the study being that behavioral data such as interaction and movement obtained through the use of wearable sensors is particularly useful for research on organizational behavior.

In a typical example for the purpose of this research, a doctor provides smartwatches to victims of convulsive seizures. The victims are located in different places where but the doctor may not be able to respond quickly one. When any patient has a seizure upon , the incidence of a seizure attack of any patient, the watch detects the repetitive motion and sends out signals to caregivers nearby for immediate action, without having to be alerted by any additional input from the doctor.

**WEARWRITE**

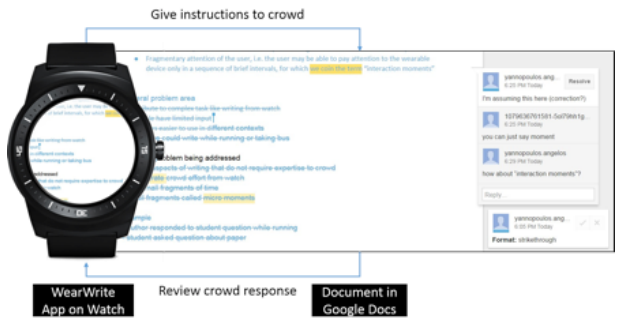

**Methodology**

Two preliminary experiments assessed the viability of the design. The experiments involved hiring four to five hourly writers to work on a document shared using Google Docs. Google Docs allows any number of workers to contribute to a document at the same time and also includes a comment system. Each experiment asked workers to write a paper. While the researchers did not provide the initial instructions or feedback and comments from the watch, their interactions with workers were constrained to limitations imposed by watch interactions.

The first preliminary trial instructed crowdworkers hired using Amazon's Mechanical Turk to shorten a biography. The results from the survey were compiled into a spreadsheet and given to writers hired using the freelancing platform oDesk.com; these workers were asked to write a research paper. Those hired were also asked to create images, graphs and charts to support the document. While this method produced high quality local content, it produced less high-level framing content. Worker writing quality varied as did the quality and type of images provided by oDesk hires. The nature of Google Doc's editing system resulted in an overwhelming number of edits that were difficult to approve or reject individually, and it seemed that workers desired more structure and feedback.

The second preliminary trial provided workers hired on oDesk.com with a document template and bulleted suggestions for each section of the report; workers were also given images to use in the document. The content for the second trial was based on a previously published paper; only one worker found the original source. In this trial, workers frequently commented on each other's work. While collaboration between workers was high, the notifications system in Google Docs lacked necessary context for researchers to provide comments and feedback as efficiently as possible.

**WEARWRITE**

We now describe the WearWrite system. The system was developed iteratively. After discussing our experiences with the first few attempts and sharing what we learned, we present the details of the final system.

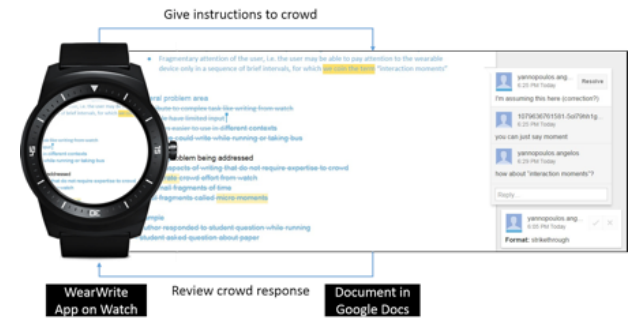

**Iterative Design**

To develop the WearWrite system, we iteratively explored two psuedo-watch-based experiences for writing academic content. Each experiments assessed the viability of the design. The experiments involved hiring four to five hourly writers to work on a document shared using Google Docs (drive.google.com). Google Docs allows any number of workers to contribute to a document at the same time and also includes a commenting system. Each experiment asked workers to write a paper. In these preliminary experiments, the researchers did not provide the initial instructions, feedback, or comments from the watch. However, their interactions with workers were constrained to limitations imposed by watch interactions.

The first preliminary trial involved writing a 2-page paper based on data collected by the authors, that was eventually titled *A Survey of Shortening Tasks in Crowdsourcing Markets*. To collect the data in the paper, the authors hired crowd workers using Amazon's Mechanical Turk, asked them to shorten a snippet of text, and asked a few follow up survey questions. The data were compiled into a spreadsheet and given, along with a few bullets highlighting key findings, to workers hired using the freelancing platform oDesk.com. Workers were asked to use this content to write a research paper, creating the necessary graphics to support the content. The authors approved or rejected each change. While this method produced high quality local content, it produced limited framing content. Worker writing quality varied as did the quality and type of images provided by oDesk hires, and workers wanted more structure. The nature of Google Doc's editing system resulted in an overwhelming number of edits that were difficult for the authors to approve or reject individually.

paper was written by workers hired on oDesk.com with work overseen by the research team from smartwatches.

**Watch Interaction**

The experimental paper-authoring software developed enables a smartwatch user to orchestrate crowd workers by supporting editorial interactions. First of all, it supports direct but limited authoring of new content via speech recognition. These input mechanisms are also used to communicate with crowd workers by replying to their comments. Second, basic formatting operations are directly enabled; for the experiment only sections, paragraphs and bullet lists were needed. Third, an editing workflow was supported. As crowd workers edit the document, their changes are handled as "suggested edits". These suggested edits are extracted by and displayed on the smartwatch. The expert receives only the actual edits made by the crowd workers as text, and a screenshot of the entire page in which an edit was made is provided as metadata for context. The expert can accept or reject the suggested edits. The expert also sees the document comments written by crowd workers, and is able to respond in follow-up comments. The whole document can be browsed or a thumbnail representation of the document's pages can be viewed. Finally, suggested edits and comments appear on the smartwatch together with relevant context, to facilitate the expert.

**Architecture and Implementation**

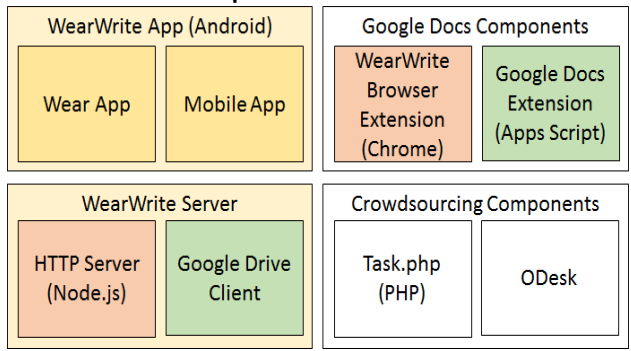

WearWrite's architecture can be broken down in four major interconnected blocks. On the mobile side, we have WearWrite actual android app with a WearWrite Server. In the Web Browsers area, it implements an extension for Google Chrome and Google Docs each that coordinate with HTTP Server in WearWrite Server and Google Drive Client respectively. For the purposes of crowdsourcing, workers are hired from ODesk and the tasks are assigned using a PHP Script.

1. **WearWrite App**: consisting of an Android Mobile app running on the phone and an Android Wear app running on the watch
2. **WearWrite Browser Extension**: used for taking screenshots, extracting edits/comments from the document, and triggering the insertion of new edits

**Watch Interaction**

WearWrite enables a smartwatch user to orchestrate crowd workers by supporting editorial interactions. First, it supports direct but limited authoring of new content via speech recognition. Authors can add sections, paragraphs, and bullets. Second, as crowd workers edit the document, their changes are displayed on the smartwatch where the expert can approve or reject them. Edits are sent along with a screenshot of the entire page in which an edit was made. Finally, comments written by crowd workers are sent to the smartwatch and the author can reply using speech. The whole document can be browsed in a thumbnail representation.

**Architecture and Implementation**

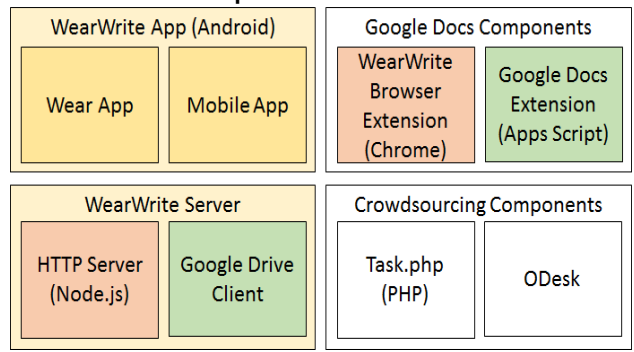

Figure 1: The four main components of the WearWrite system, which allow crowd workers on desktop interfaces and an author on a smartwatch to collaborate.

WearWrite's architecture can be broken down into four major interconnected blocks. On the author side, a WearWrite Android application communicated with an Android Wear smartwatch and also the WearWrite Server. The WearWrite server connects to Google Docs to make changes, notice changes or comments, and facilitate discussion. Crowd workers are recruited from oDesk, and coordinated using a task queue that contains a number of generic paper-writing tasks, e.g., "turn the bullets in the 'Introduction' into paragraphs," "find the paper that each bullet in the 'Related Work' section refers to and create a reference to it."

via Google Docs Extension and new comments via Google Drive Client into the document[2]
3. **Google Docs Extension**: used for analysing document structure of, and insertion of new edits into, the document
4. **WearWrite Server**: 1) HTTP server used by a) the WearWrite App for posting new edits/comments from the watch to the document and b) the WearWrite Browser Extension for uploading screenshots of the document; and 2) Google Drive Client used for insertion of new comments into the document
5. **Task.php**: directs workers to specific tasks to provide lightweight structure
6. **ODesk**: used for recruiting workers

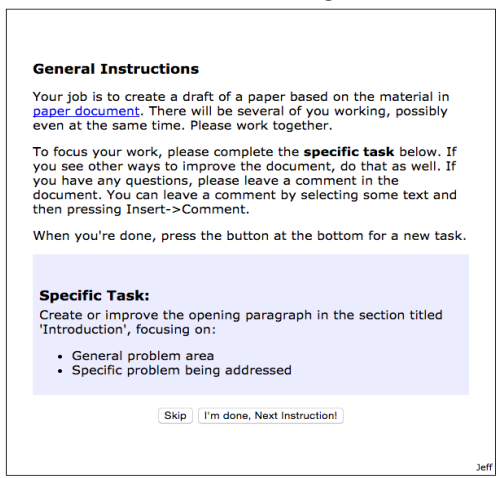

**MAIN EXPERIMENT**

The main experiment performed to validate the paradigm proposed in this paper implemented the custom prototype described above and used it to write the first draft of this paper. Described in the previous section was the prototype's architecture and implementation and an analysis of the experiment's results appears in the following section. This section focuses on the method, followed by the experiment itself, particular in relation to the collaborative authoring process.

Two scenarios, an ideal scenario and a practical scenario, were defined for the experiment. The ideal scenario exactly corresponds to the paper's proposed paradigm. The practical scenario includes certain compromises which were made so that the experiment could be conducted with reasonable resources. Development time

---
[2] at the time of writing, insertion of comments is only possible via Google Drive SDK and not via Google Apps Script, so we need both components to support both actions

available for the software prototype constituted the largest resource limitation, so the practical scenario was followed in the experiment.

In the ideal scenario, the watch user performs all interactions using exclusively the smartwatch. The watch user can receive general-purpose emails on the smartwatch, but exclusively uses the WearWrite prototype, running on the smartwatch, for all crowd sourcing and document writing activities. The workflow followed is:

- The watch user receives a request for information through general-purpose email on the smartwatch.
- The watch user recruits a number of crowd workers through the smartwatch. In applications where crowd workers are available on short notice, it is assumed that a preparatory recruitment process has already been performed using a desktop computer, and that in this scenario the watch user communicates using the smartwatch with crowd workers who are on standby and recruits those who are available when needed.
- After sending requests to crowd workers, the watch user efficiently uses time spent waiting for replies to seed the document. The seed document is created with Google Docs and authored exclusively from the smartwatch. It contains the basic structure of the document, the key points it should address, and references for crowd workers to use.
- When crowd workers become available, an iterative authoring process begins. The crowd workers transform the seed document into the desired final document, step by step. They work on concurrently through Google Docs, and every edit they make is handled as a suggestion. The suggestion is sent to the watch user to accept or reject, and potentially comment on. Also, any comments made by crowd writers are sent to the watch user, who can reply to them if necessary. The watch user can also view the entire document on the smartwatch, although this is not easy on the small watch display. At the end of this process, the document draft is complete.

The practical scenario differs from the ideal scenario in the following ways:

- The task of recruiting crowd workers was performed from a desktop computer
- Overcoming the occasional prototype bugs was achieved by using a desktop computer if necessary. However, the desktop was only used to perform actions that would normally have been performed using a bug-free prototype.
- The watch user spent 99% of his time using the smartwatch, not a desktop computer
- On a few occasions, crowd workers emailed the watch user. In these cases, they were asked to repeat their message through the google doc.

**WRITING THIS PAPER**

We now turn to our experience writing this paper with WearWrite. Our original intention was to write *everything* on the watch, but this proved unwieldy. In practice, we expect authors will switch between the watch, a mobile, and a desktop computer, as is convenient for them.

In the ideal scenario, the watch user performs all interactions using exclusively the smartwatch. The watch user can receive general-purpose emails on the smartwatch, but exclusively uses the WearWrite prototype, running on the smartwatch, for all crowd sourcing and document writing activities. The workflow followed is:

1. The watch user receives a request for information through general-purpose email on the smartwatch.
2. The watch user recruits a number of crowd workers through the smartwatch. In applications where crowd workers are available on short notice, it is assumed that a preparatory recruitment process has already been performed using a desktop computer, and that in this scenario the watch user communicates using the smartwatch with crowd workers who are on standby and recruits those who are available when needed.
3. After sending requests to crowd workers, the watch user efficiently uses time spent waiting for replies to seed the document. The seed document is created with Google Docs and authored exclusively from the smartwatch. It contains the basic structure of the document, the key points it should address, and references for crowd workers to use.
4. When crowd workers become available, an iterative authoring process begins. The crowd workers transform the seed document into the desired final document, step by step. They work on concurrently through Google Docs, and every edit they make is handled as a suggestion. The suggestion is sent to the watch user to accept or reject, and potentially comment on. Also, any comments made by crowd writers are sent to the watch user, who can reply to them if necessary. The watch user can also view the entire document on the smartwatch, although this is not easy on the small watch display. At the end of this process, the document draft is complete.

In practice, we compromised in this a bit:

1. Crowd workers were recruited from the oDesk desktop computer web interface.
2. Overcoming the occasional prototype bugs was achieved by using a desktop computer if necessary. However, the desktop was only used to perform actions that would normally have been performed using a bug-free prototype.

The practical scenario is equivalent to the ideal scenario for the purposes of this experiment because the collaborative authoring process was executed entirely through the watch interface as intended; rarely using a desktop computer to overcome a prototype bug, but even then only performing actions as the prototype was designed to do. The recruiting process is an important part of the practical crowdsourcing activity, but it is not a part of the collaborative authoring process which is the focus of the experiment. Similarly, sharing the watch user role between researchers did not impact the interface usage or the collaborative workflow which were the factors being researched.

**RESULTS**

**Subjective Assessment**
In the course of the third and main experiment, the interactions between crowd workers and watch users were systematically observed. Our observations and conclusions can be classified as related to a) preparation and task allocation, b) basic effectiveness, c) communication, d) difficulties, e) involvement and creativity. We discuss each category below.

Preparation and task allocation. The first interactions between the crowd workers and the watch user were through a third party website, oDesk.com. The normal process of recruitment concluded with crowd workers receiving their initial instructions. Once they started work on the collaborative authoring job itself, the crowd workers exhibited a constructive pattern of behaviour. They started by working through the entire seed document, in order to understand the research they were documenting. Next, they worked on random tasks allocated to them by the Task.php webpage. The randomisation was successful in allowing multiple crowd workers to collaborate on the authoring job without prior coordination. Some small difficulties did arise due to the randomisation, however, as discussed below. For instance, a crowd worker was asked by Task.php to write the paper's abstract before the rest of the paper had been composed.. As the crowd workers became more accustomed to the collaboration dynamic in the experiment, they took increasingly greater initiative in simply choosing the section of the paper they felt they were best able to contribute to.

*Basic effectiveness*
The crowd workers were, in general, effective at producing content. The entire paper was written within three days. The quality of content produced was found by the watch user to be sufficiently high. A metric that indicates this is the watch user's significant "idle time" during the authoring process: this means that the watch user had time during which to make corrections or improvements to the document, but did not need to do so, since it was found to be already of satisfactory quality.

3. The watch user spent nearly all of his time using the smartwatch, not a desktop computer.
4. On a few occasions, crowd workers emailed the watch user. In these cases, they were asked to repeat their message through the Google Doc, where they were responded to using WearWrite.
5. Two different researchers adopted the role of watch user at various times.
6. Finally, in the actual experiment, five crowd workers were recruited

**EXPERIENCES WRITING A PAPER ON A WATCH**
In this section, we report on writing this paper.

The interactions between crowd workers and WearWrite users were systematically observed while writing this paper and during the iterative design process. Our observations and conclusions can be classified as related to *a)* preparation and task allocation, *b)* basic effectiveness, *c)* communication, *d)* difficulties, *e)* involvement and creativity.

*Preparation and task allocation*

The first interactions between the crowd workers and the watch user were through a third party website, oDesk.com. The normal process of recruitment concluded with crowd workers receiving their initial instructions. Once they started work on the collaborative authoring job itself, the crowd workers exhibited a constructive pattern of behaviour. They started by working through the entire seed document, in order to understand the research they were documenting. Next, they worked on tasks allocated to them by *task.php*. Randomized task allocation allowed multiple crowd workers to collaborate on the authoring job without prior coordination. However, some problems occurred. For instance, a worker was asked to write the paper's abstract before the rest of the paper had been composed. As the crowd workers became more accustomed to the collaboration dynamic in the experiment, they took increasingly greater initiative in simply choosing the section of the paper they felt they were best able to contribute to.

*Basic effectiveness*
Workers were, in general, effective at producing content. The entire paper was written within three days. Readers

The watch user could see how the document progressed, and it was sufficient to occasionally provide the crowd workers with suggestions and feedback on edits.

The crowd workers were comfortable working from the original bullet points comprising the seed document. They used these bullet lists both as content expressing the paper's ideas, and as a to-do list from which they could work. They progressed through the work systematically, and marked ongoing and completed work for each other (for example, using a strikethrough effect on completed bullet points, but not deleting them).

Furthermore, the crowd workers performed intelligent and critical work. Beyond their basic task of transforming notes into well structured text, they made suggestions and improvements ranging from re-wording keywords and key phrases in the notes to planning and executing ideas for the presentation.They made suggestions about the structure of the document as well as comments that were valuable at the research level, such as how best to evaluate the experimental results. The more creative contributions of the crowd workers are discussed below.

*Communication*
The crowd workers approached their task as a highly collaborative process. They regularly made comments to and requested feedback from the watch user. They used document comments to request clarifications from the watch user if they felt the instructions, the seed document, or another crowd worker's text to be unclear. In turn, sometimes text or comments written by the crowd workers were vague, in which case the watch user initiated a discussion in the document comments. The discussion was rapid, concrete and constructive, as as evidenced by the fact that the entire document was completed within three days.

Although communication through document comments was performed successfully, some crowd workers were unsatisfied with the approach, suggesting, for example, in-document "chat" which is not supported through the watch. Future work could certainly provide useful improvements by offering a better discussion interface.

*Difficulties*
The initial learning curve faced by crowd workers is steep enough to cause some initial problems and challenges. While these problems and challenges are by no means particularly serious, they do force the watch user to supervise the authoring process closely if a job is to be completed efficiently. One crowd worker, following the randomised task assignment process of Task.php, started to write the abstract before working on the rest of the document, missing the key points. The watch user intervened by making a comment on the half-written abstract and deferring its completion to a more appropriate time.

can judge for themselves, but the authors were surprised by the quality of the content produced. The authors had significant "idle time" during the authoring process. The author could see how the document progressed, and it was sufficient to occasionally provide the crowd workers with suggestions and feedback on edits.

The crowd workers appeared comfortable working from the original bullet points comprising the seed document. They used these bullet lists both as content expressing the paper's ideas, and as a to-do list from which they could work. They progressed through the work systematically, and marked ongoing and completed work for each other (for example, using a strikethrough effect on completed bullet points, but not deleting them).

Furthermore, the workers performed intelligent and critical work. Beyond their basic task of transforming notes into well-structured text, they made suggestions and improvements ranging from re-wording keywords and key phrases in the notes to planning and executing ideas for the presentation. They made suggestions about the structure of the document as well as comments that were valuable at the research level, such as how best to evaluate the experimental results. The more creative contributions of the crowd workers are discussed below.

*Communication*
The crowd workers approached their task as a highly collaborative process. They regularly made comments to and requested feedback from the author. They used document comments to request clarifications from the author if they felt the instructions, the seed document, or another crowd worker's text to be unclear. In turn, sometimes text or comments written by the crowd workers were vague, in which case the watch user initiated a discussion in the document comments. The discussion was rapid, concrete and constructive, as evidenced by the fact that the entire document was completed within three days.

Although communication through document comments was performed successfully, some crowd workers were unsatisfied with the approach, suggesting, for example, in-document "chat" which is not supported through the watch. Future work could certainly provide useful improvements by offering a better discussion interface.

*Difficulties*
Workers experienced a bit of a learning curve before they became comfortable with WearWrite. One crowd worker, following the randomised task assignment process of task.php started to write the abstract before working on the rest of the document, and missed many of the key points. The watch user intervened by making a comment on the half-written abstract, requesting that it be deferred until later. Authors may sometimes start with abstracts, but this is difficult for workers new to the project to do without necessary context.

In general, linear progression by crowd workers through the document is not a good idea. It is best for the task allocation process to make clear to crowd workers that task suggestions can be skipped until they are comfortable with a particular task. Beyond this, moving from fully randomised task allocation to a state where crowd workers are comfortable enough to choose tasks for themselves is advantageous, since crowd workers will perform better in this case. However, some randomisation and/or centralised control of the task allocation process needs to be maintained to avoid a case where crowd workers "pick low hanging fruit" and certain writing tasks are left unperformed because nobody volunteered for them.

Additional difficulties include maintaining a cohesive style and terminology (expert, domain expert, watch user, smartwatch user, main author were used interchangeably by the various crowd workers),managing the length of the document and in particular avoiding the word count increasing too much, and complicated change management, which was not needed for this experiment, though it is expected to be difficult if required.

Some writing tasks that would be easy for the watch user can be unexpectedly difficult for the crowd workers. For example, the simple task of summarising related work once specific and sufficient references have already been identified can be difficult for crowd workers. They are not experts in the field, and it is hard for them to confidently speak in critical terms about research they are not familiar with.

Finally, the watch user also faced certain difficulties. The speech recognition on the watch produced transcription errors that were sometimes confusing. Also, there were some editing bugs that were hard to explain during the experiment; on one occasion, the watch user attempted to accept an edit made by a crowd worker but the edit was deleted instead. However, the fact that the paper was successfully authored in a brief period of time using only the intended interfaces  indicates these difficulties are minor..

*Involvement and creativity*
The crowd workers took ownership of the writing task and demonstrated significant amounts of involvement and creativity. There was a suggestion for an alternative scenario for the paper's proposed paradigm, which is too complex to base an experiment on, but functions as a good, realistic motivating scenario. Feedback was given on terminology defined in the seed document, which was in some cases accepted and improved the document's clarity, for example "interaction moments" replaced the seed document's term "mico-moments", or the term "watch metaphor", which was discarded.

The crowd workers also made long-term contributions. They communicated their reflections on task design and

In general, linear progression by crowd workers through the document is not a good idea. The task allocation process was altered for this final version to make clear that workers could skip its suggestions. Beyond this, moving from fully randomised task allocation to a state where crowd workers are comfortable enough to choose tasks for themselves is advantageous, since crowd workers will perform better in this case. However, some randomisation and/or centralised control of the task allocation process needs to be maintained to avoid a case where crowd workers "pick low hanging fruit" and certain writing tasks are left unperformed because nobody volunteered for them.

Additional difficulties include maintaining a cohesive style and terminology (expert, domain expert, watch user, smartwatch user, main author were used interchangeably by the various crowd workers – as we do this final editing pass we are trying to converge on vocabulary, but future systems may try to get workers to agree on these terms before contributing too much text), managing the length of the document (indeed, this document grew to be too long), and complicated change management, which was not needed for this experiment.

Some writing tasks that would be easy for the author can be unexpectedly difficult for workers who lack specific expertise and context. For example, the "Related Work" section ended up the poorest of all of the document, despite the author having provided projects to reference and snippets describing their importance. Workers are not experts in the field, and it is difficult to write critically about unfamiliar research.

Finally, the author faced certain difficulties using WearWrite. The speech recognition on the watch produced transcription errors that were sometimes confusing. Also, there were some editing bugs that were hard to explain during the experiment; on one occasion, the author attempted to accept an edit made by a crowd worker but the edit was deleted instead. However, the fact that the paper was successfully authored in a brief period of time using only the intended interfaces indicates these difficulties are relatively minor.

*Involvement and creativity*
The crowd workers took ownership of the writing task and demonstrated significant amounts of involvement and creativity. There was a suggestion for an alternative scenario for the paper's proposed paradigm, although we found it to be too complex. Feedback was given on terminology defined in the seed document, which was in some cases accepted and improved the document's clarity, for example "interaction moments" replaced the seed document's term "mico-moments", or the term "watch metaphor", which was discarded.

The crowd workers also made long-term contributions. They communicated their reflections on task design and on the smartwatch can introduce confusion by incorrectly transcribing the watch user's comments. They made suggestions about how the current experiment might be more extensively assessed, for example by tracking document metrics over time/effort. They provided ideas for future research, for example integrating voice interactions with document comments based interactions.

At the same time, the crowd workers were sufficiently detached from the writing task to be able to think as peer-reviewers for the paper. This is in fact also a creative contribution, as it often starts a discussion that leads to improvements of the paper. For example, it was suggested to clearly explain why the topic for the text collaboratively authored during the experiment was representative.

**Objective Assessment**
It is important to use objective metrics to assess the experiment described in this paper. Although an extensive subjective analysis is presented above, an objective assessment is also necessary. However, since the work presented here is simultaneously an experiment and a learning process about the proposed paradigm, it was not feasible to design an extensive verification protocol before conducting the experiment. Therefore, this section describes a very simple objective assessment. and discusses designs with more complicated metrics for similar experiments in the future.

The central hypothesis of this work is that a paper of high academic standard can be produced using WearWrite and crowdsource writers. Although the process followed in the proposed paradigm is thoroughly discussed, the key question is regards the quality of the final product produced. In order to objectively assess academic papers, the commonly accepted system is independent academic peer-review. This system is imperfect, but it is generally accepted as the best available method. For this experiment, N colleagues, who were not involved in this work, were asked to review the paper and deliver an assessment according to their normal standards (accept/major revisions/minor revisions/reject) when reviewing for academic publications. The only instructions given to reviewers indicated they should not judge the paper in terms of innovation, because the authoring process is the central concern and not the originality of the proposed paradigm. The result was [...]

For future work, we suggest the following measurements for the proposed authoring process.

- Measure direct edits made by the watch user to crowd workers' text. A percentage based on word count can be calculated. A good result is under 1%, 1% to 5% is acceptable.
- Measure edits made by crowd workers to each other's text. A percentage based on word count can be calculated. A good result is between 5% and 15%.

A result up to 5% or between 15% and 25% is acceptable. Note that it is a good sign when crowd workers pay attention to each other's work, because this indicates collaboration. Too much re-writing of the same material is a signal of inefficiency, however.

- Measure document completion velocity in words per person-hour, and tracked as a function of time. Absolute values for this measure are of secondary importance, rather the question is whether velocity fluctuates excessively. Further experimentation is needed to define desired values for this measure; however, as an example, it would present a significant problem if the proposed paradigm works well for authoring 80% of the paper, but the crowd workers find the hardest 20% of the paper extremely challenging and their velocity for these sections drops to 20% of their velocity for the rest of the paper.
- A single, expert author could write a paper on a specific subject and measure the time required to do so. The paper can be re-written using crowd workers. A percentage of person-hours can be calculated for the ratio of the collaborative effort to the single author effort, and another for the ratio of the effort made by the watch user in the collaborative effort to the total collaborative effort. A good result is up to 200% and 20% respectively. A result of up to 300% and 50% is acceptable. Note that the watch user is technically unable to perform this work alone, so the paradigm is successful in spending crowdsourced effort to enable an otherwise impossible task. The goal of this project is not to reduce the total amount of effort required to write academic papers.

**DISCUSSION**

The results taken together indicate that "useful computing tasks which cannot be wholly performed using wearable devices due to their constrained interaction capabilities can be successfully partially crowdsourced from wearable devices". New software developed for wearable devices enables its users to "contribute the necessary expertise" and to "manage the contributions of crowd workers", allowing the user to author a research paper using a wearable device and collaborating with crowd workers. With this system, one achieves the same results as if working at a desktop computer.

The software and experiments presented were able to satisfactorily test the limits of wearable interface capabilities. Where the smartwatch device could not directly fulfill the user's requirements, it was instead programmed to enable the user to orchestrate crowd workers. Thus, research and authoring tasks that cannot be directly performed on a smartwatch platform were transformed into collaboration tasks. The smartwatch user provides the necessary expertise, editorial judgement, and workflow management. Crowd workers provide additional authoring capability. Since the crowd workers

**DISCUSSION**

Together, results indicate that "complex tasks that are difficult to perform on wearable devices due to their constrained interaction capabilities can still be successfully crowdsourced from wearable devices." WearWrite enables its users to "contribute the necessary expertise" and to "manage the contributions of crowd workers," allowing the user to author a research paper using a wearable device and collaborating with crowd workers. With this system, one achieves results comparable to those from a desktop computer.

This approach might be useful for completing other types of complex tasks or for completing such tasks on other types of wearable devices, such as head mounted displays. It remains an open question for future research and philosophical consideration to address questions of the influence of this paradigm on cognitive processes. How would allowing work to invade these small moments, which might otherwise be rest from intellectual work, affect psychological health? The potential negative influence of constant "uptime" performing intellectually challenging tasks may be severe.

WearWrite breaks authoring tasks into small steps. Some steps are performed by the expert, while others are outsourced to crowd workers. Even the tasks performed by the expert are performed in small "interaction moments." However, at what point does this seeming benefit become a liability, if for instance, by splitting our attention among multiple unrelated authorings, necessary context instantiation is lost? As a counterargument, is it perhaps the case that by outsourcing the less creative parts of a project, intellectually intensive work is reserved for the situations in which it is actually merited.

perform a genuinely creative task, they need to understand the expert's guidance and genuinely collaborate, i.e. they are not "mindless" or "drone" workers. This is why the new platform needs to provide genuine collaboration tools; for example, fast interaction through comments, fast acceptance or rejection of suggested edits, and editorial selection between several crowd-offered suggested edits to the same text.. Once real collaboration is enabled, the crowd workers are able to perform authoring work that the expert user is unable to perform on the wearable device.

It is particularly interesting to explore use of this approach for other wearable devices, such as head mounted displays. This work shows that it is technically possible to enable the described paradigm. It remains an open question for future research and philosophical consideration to address questions of the influence of this paradigm on cognitive processes.

The paradigm helps break an authoring task into small steps. Some steps are performed by the expert, while others are outsourced. Even the tasks performed by the expert are performed in small "interaction moments". Although technically this paradigm enables a desirable level of quality, is it harder to achieve from a cognitive point of view?

Furthermore, even if work is high quality, how would allowing work to invade these small moments, which would otherwise be rest from intellectual work, affect psychological health?? The potential negative influence of constant "uptime" performing intellectually challenging tasks may be severe.

Another question is that of context. The proposed paradigm enables multitasking, a powerful capability. However, at what point does this seeming benefit become a liability, if for instance, by splitting our attention among multiple unrelated authorings, necessary context instantiation is lost? As a counterargument, is it perhaps the case that by outsourcing the less creative parts of a project, intellectually intensive work is reserved for the situations in which it is actually merited.

Breaking up and carefully orchestrating the authoring process grants an opportunity to study the process itself. For example, the collaborative writing process could offer insight on the learning process in order to help students improve their writing.

Finally, we must also discuss the role of crowd workers in this paradigm. As stated above, it is technically possible to enable the described paradigm. However, once the collaborative tools are in place, the value contributed by each collaborator needs to be assessed. Should crowd workers be considered amongst the authors of the paper or other written work? In this context, it should be noted that many co-authorship scenarios exist where collaborators are not, in practice, given credit: for

WearWrite breaks authoring tasks into small steps. Some steps are performed by the expert, while others are outsourced to crowd workers. Even the tasks performed by the expert are performed in small "interaction moments." However, at what point does this seeming benefit become a liability, if for instance, by splitting our attention among multiple unrelated authorings, necessary context instantiation is lost? As a counterargument, is it perhaps the case that by outsourcing the less creative parts of a project, intellectually intensive work is reserved for the situations in which it is actually merited.

Breaking up and carefully orchestrating the authoring process grants an opportunity to study the process itself. For example, the collaborative writing process could offer insight on the learning process in order to help students improve their writing.

Finally, we must also discuss the role of crowd workers in this paradigm. Should crowd workers be considered authors of papers written in this way? In this context, it should be noted that many co-authorship scenarios exist where collaborators are not, in practice, given credit: for example, grammar editors, ghostwriters, and technical writers, etc. We have listed the worker who contributed to this paper as authors of it. The contributions of the crowd workers are harder to encapsulate in a role as simple and atomic as these. As we mentioned earlier, workers often participated creatively in the writing process.

example, grammar editors, ghostwriters, and technical writers, etc. However, in the paradigm presented in this paper, the contributions of the crowd workers are harder to encapsulate in a role as simple as these. An interesting extension of this work would be to facilitate collaboration between professionally related individuals, for example allowing senior research lab members to continuously mentor and guide the more junior members: here, the authorship question is particularly relevant.

**CONCLUSION**

This paper introduced a new paradigm for orchestrating crowd workers from a wearable device, with the result of transcending the limitation of the wearable device's interfaces. It enables a watch user, who is an expert in a particular domain, to contribute knowledge and guidance, so as to complete a knowledge-intensive task that requires her expertise. It also enables the watch user to leverage the capabilities of the crowd so as to perform tasks that are impossible, or extremely inefficient, to perform using the wearable device.

We conducted an experiment that practically demonstrated how an academic paper can be authored using the proposed approach. Many promising conclusions were drawn from the analysis of the experiments results: limited communication was sufficient even for the knowledge intensive task that was pursued; limited expert guidance helped the crowd workers not only to perform high-quality basic work, such as constructing good prose from the expert's brief notes, but it also enabled them to contribute creatively and critically to the task; and the difficulties that arose during the experiment were surmountable and can be effectively handled.

The proposed paradigm is expected to lead to other, new systems where domain expertise is injected from a user of a wearable device, and/or workflow management is provided through a wearable device. It should also be noted, though, that interesting questions arise regarding the impact of wearable devices on the way work is performed, and on the people who perform it, and some important examples of this were discussed.

**CONCLUSION**

We have introduced WearWrite, an application for Android Wear that allows authors to write academic papers from their smartwatches. We believe this system represents a new paradigm that may change how and where we work together to get work done. Completing complex tasks within the constraints of wearable devices brings up a host of interesting questions regarding the impact of wearable devices on the way work is performed, and on the people who perform it that we have only started to address.

*The following is the initial outline that we gave workers. Each section and bullet point were created with WearWrite. Speech recognition errors made this somewhat frustrating, as much of the input contained errors. Some words were not in the speech recognizer's vocabulary, e.g., WearWrite. We fixed these manually as shown below. We imagine a released system may send the audio of the author's instruction along with the text so that the crowd can do the fixes. In addition to the bullets of text, we also provided the workers with the images and figures to be used in the paper. It is interesting to consider whether the crowd could also help authors create images and figures from the watch, but we did not explore that in this paper.*

# Title

- WearWrite: Writing and Editing Papers from a Watch
- WearWrite: Orchestrating the Crowd to Complete Complex Tasks from Wearables
- We Wrote This Paper On A Watch

**ABSTRACT**

**INTRODUCTION**

**general problem area**
- contribute to complex task like writing from watch
- wearable ~~table~~have limited input ~~new line~~
- wearables easier to use in different contexts ~~new line worry~~
- what if ~~he~~we could ~~ride my~~write while running ~~on~~or taking bus

**Problem being addressed**
- offload aspects of writing that do not require expertise to crowd
- ~~orchestrade~~orchestrate crowd effort from watch
- ~~used~~use small fragment of time
- small fragment called micro moments

**Example**
- author ~~is one of the Student~~responded to student question ~~why~~while running
- ~~ask~~student asked question about paper
- ~~autoresponder~~responded via voice without stopping
- ~~injector~~injected expertise quickly

**Thesis statement**
- input on ~~where abouts~~wearables is constrained
- complex ~~task difficulty difficulty~~tasks difficult to complete
- maybe able to orchestrate ~~alright are these~~ or direct others to complete complex tasks
- it is possible to manage the process of writing a paper from a watch

**Contributions**
- motivated directing writing from a watch
- introduce system for writing paper from a watch
- present case study ~~dorada~~to validate the approach
- offer ~~in size~~insights for future work in this area

**RELATED WORK**
- work related to crowdsourcing
- work related writing with crowd
- related to ~~variable~~wearable interaction

**Crowd writing**
- crowd ~~were closed~~workflows already support complex tasks
- ~~silent~~bernstein's soylent: crowds ~~and better than~~embedded in microsoft word; find fix verify pattern
- kittur's ~~cloudforge cloudword~~ crowdforge: crowd work management framework
- shepherd: idea of promoting ~~health workers~~crowdworkers to leaders who provide feedback
- ~~example~~ bernstein's ensemble: creative writing with a crowd
- automated writing support tools
- platforms with feedback for writers ~~for example Bluhm used and reflective~~
- turkit's parallel and serial writing with a crowd
- self sourcing
- quickly capture idea with the ~~watchmen~~watch when writing is inconvenient

**Watch interaction**
- watch interaction is limited
- two strategies explored for ~~riding~~writing on watch so far
- improved hardware for input such as harrison's smart watch face
- designing for text entry right on the watch
- zoom board ~~fence whiteboard~~and swipeboard
- in contrast we want to use crowdsourcing to overcome small device limitations

**WEARWRITE**

**Initial experiments**
- two preliminary experiments to inform design of way right
- recruited for ~~25~~4-5 workers to shared google doc

- workers paid hourly
- in each we asked workers to write a paper
- we did not write these papers from the ~~watchmen~~watch
- ~~but~~ constrained our interaction to be similar to watch interactions
- first ~~trial was on~~ experiment ~~that~~ had mturk workers shorten a document
- gave data to odesk workers and ~~ask~~asked them to write a paper
- this led to several interesting insights
- produced high quality local content
- produced less high-level framing content
- ~~work equality~~worker quality varied
- workers produced different ~~call the~~quality and type of images
- ~~improving~~approving every edit was too much
- seemed like workers needed more structure

- second ~~try again~~trial gave workers document template with bullets
- gave workers images
- contents from a previously published paper
- google ~~dot~~doc notifications ~~like~~lacked some necessary context
- often workers would ~~come into~~comment to one another
- one ~~work~~worker found the original paper this project was based on

**Watch UI**
- add screenshot

**Supported interactions**
- add content to google docs from watch
- supports sections paragraphs bullet points
- receive suggested edits on watch
- receive ~~commands~~comments or replies on watch and reply from watch
- view thumbnail representation of pages
- gives context to edit or comments as they appear on watch

**Architecture and implementation**
- ~~to do inside~~add figure ~~here~~
- describe main components

**EXPERIMENT**
- writing a first ~~draught of this paper without using prototype writing a first draught~~draft of this paper with a crowd using ~~a~~our prototype

**Method**
- created ~~see~~seed document containing basic structure key points and references from watch
- made minor edits to seed document after writing from watch
- recruited 5 crowd workers on ~~a desk~~odesk
- interactions mostly ~~perform~~performed from the watch

- ~~no~~new results were added to the document as they came in

**RESULTS**
- ~~to do~~add results~~go here~~

**DISCUSSION**

**Watch as form factor**
- prototype pushes ~~watermill~~watch metaphor as much as possible
- using crowd ~~make~~makes some tasks watch friendly that wouldn't be otherwise
- ~~tooting text~~choosing text from several crowd offered suggestions is relatively easy
- tasks that are not watch friendly may be done on another device
- the question ~~that~~then is how to support switching
- ~~ar~~our techniques likely ~~also~~apply to other ~~variables~~wearables such as google glass

**Writing in micro tasks**
- do we lose value from what we do when we break it down into small ~~pines~~points
- even if we have micro moments to use how does it impact us to fill them with ~~cock in headly~~cognitively demanding tasks
- when people are writing ~~Mather~~multiple papers context ~~is insulation~~ instantiation maybe more important
- does farming of small pieces and ~~louis~~allow us to focus on the big picture when we have a chance
- maybe you can learn from the process is this the way to educate students about how to write papers

**Using hard workers**
- when are workers authors
- there are many ~~quarters ships~~co-authorship scenarios where people do not get credit
- english grammar editors ghostwriters technical writers etc.~~is Hutter~~
- crowd ~~cookie jars~~could be generalized to research lab members or even ~~sing~~single author at different times ~~of~~or stages of writing process

**CONCLUSION**
- introduced approach earth orchestrating crowd from a wearable device
- possible to write a paper from the ~~watchmen~~watch
- ~~maybe~~may lead to other systems ~~wreck studies~~where expertise is injected from wearable devices
- introduces questions about how wearables may impact work

**REFERENCES**

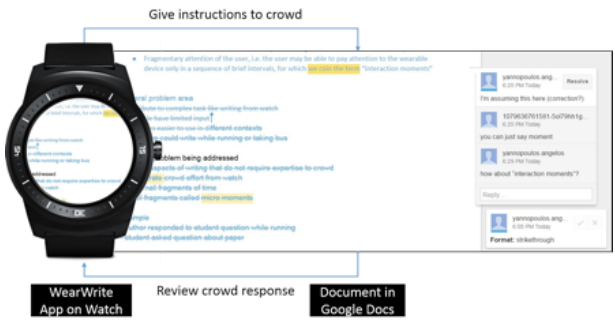

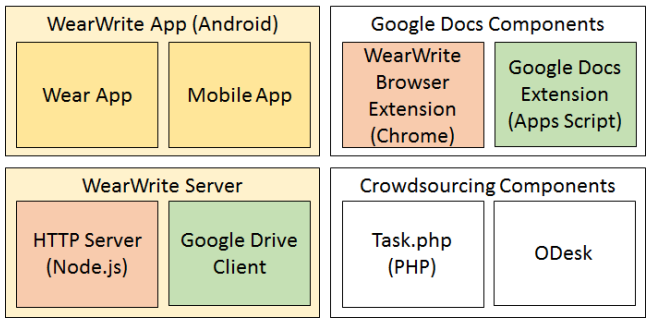